\documentclass[final,twoside,12pt]{elsarticle} %NIKOS: FOR THE SUBMISSION

\usepackage{graphicx}
\usepackage{hyperref}
\hypersetup{colorlinks=true,citecolor=blue}

\usepackage{amsthm}

\biboptions{square,numbers,sort&compress,comma}

\usepackage{units}

\journal{Proceedings of the Combustion Institute 33}

\begin{document}

%Trick for the separate title page.

\begin{frontmatter}


\title{A turbulent premixed flame on fractal-grid generated turbulence}



%%One way
%\author[mech]{CorrAuthor\corref{cor1}}
%\ead{x.xxx@imperial.ac.uk}
%
%\author[mech]{Author2}
%%\ead{x.xxx@imperial.ac.uk}
%
%\author[aero]{Author3}
%%\ead{x.xxx@imperial.ac.uk}
%
%\address[mech]{Department of Mechanical Engineering, Imperial College London, South Kensington Campus, London, SW7 2BX, UK}
%\address[aero]{Department of Aeronautics, Imperial College London, South Kensington Campus, London, SW7 2AZ, UK}
%
%\cortext[cor1]{Corresponding author. Fax: +44 20 758.}
%%One way

%%Alternative
\author{N. Soulopoulos\corref{cor1}}
\ead{ns6@imperial.ac.uk}
\author{J. Kerl\corref{}}
\author{F. Beyrau\corref{}}
\author{Y. Hardalupas\corref{}}
\author{A.M.K.P. Taylor\corref{}}
\address{Mechanical Engineering Department, Imperial College London, SW7 2BX, London, UK}

\author{J.C. Vassilicos\corref{}}
%\ead{x.xxx@imperial.ac.uk}
\address{Aeronautics Department, Imperial College London, SW7 2AZ, London, UK}

\cortext[cor1]{Corresponding author. Fax: +44 20 7594 1472.}
%%Alternative

\begin{abstract}
A space-filling, low blockage fractal grid is used as a novel turbulence generator in a premixed turbulent combustion experiment. In contrast to the power law decay of a standard turbulence grid, the downstream turbulence intensity of the fractal grid increases until it reaches a peak at some distance from the grid before it finally decays. The effective mesh size and the solidity are the same as those of a standard square mesh grid with which it is compared. It is found that, for the same flow rate and stoichiometry, the fractal generated turbulence enhances the burning rate and causes the flame to further increase its area. Using a flame fractal model, an attempt is made to highlight differences between the flames established at the two different turbulent fields.
\end{abstract}


\end{frontmatter}

% \linenumbers

\section{Introduction}
Lean premixed combustion of gaseous fuels is currently one of the most important technologies to achieve low pollutant emissions at high efficiencies in the power generation sector, \textit{e.g.} \cite{lean_combustion_gas_turbines_mcdonnel_2008}. Reduction of NO$_X$ emissions is a direct outcome of the lower combustion temperatures when burning a lean mixture and complete combustion prevents the creation of unburned hydrocarbons and carbon monoxide. The thermal efficiency advantage provides, also, the added benefit of smaller CO$_2$ emissions.

However, for example in gas turbines, flame stability and flashback are problematic areas, among others \cite{driscoll_mongia_flashback_2009, candel_flashback_2004}. Flame stability is related to, \textit{e.g.}, flame extinction as the lean limit is approached and flashback can be a problem at low heat release rates. In both cases the turbulent flame speed is a determining parameter and, in general, higher turbulent velocities are preferred. The turbulent flame speed is, for any given fuel, determined by the turbulent fluctuations of the flow \cite{bradley_turbulent_velocity_1981}, so controlling the local turbulence level is highly desirable in a variety of situations. In laboratory premixed flames a variety of square mesh grid or perforated plate designs are used at some position upstream of the flame stabilization region to generate turbulence \cite{cheng_flame_grid_turbulence_1981}. Very near the grid high turbulence levels can be achieved, which, though, die out fast following a power law decay, so the flame is stabilized a few mesh lengths downstream of the grid.

In the present paper, we propose the use of a rather new turbulence generator, a so called fractal grid to generate the turbulence in a premixed flame experiment. Fractal grids are not only very interesting for fundamental turbulence research \cite{vassilicos_fractal_prl_2010}, but are potentially, also, of great practical importance for industrial applications \cite{vassilicos_fractal_mixing_effective_2009}. Conceptually, a fractal grid differs from a standard grid because it creates turbulence by exciting many length scales of various sizes simultaneously, rather than a single one, so the underlying mechanism that generates turbulent fluctuations is different for both types of grids. The grid we use possesses a space-filling property that relates to the fractal dimension, $D_f$, of the grid having the value 2 and is elaborated upon in \cite{vassilicos_hurst_fractal_2007}, where it was also shown that the space-filling grid achieves flow homogeneity faster than fractal grids with other dimensions $D_f$. In the present work, we use  a grid that was "tailored" to generate the maximum of the turbulence intensity just at the position of the flame stabilizing wire (however, this maximum can be generated at any practical distance by using different grid designs).

The structure of the paper is as follows: section \ref{experiment}, gives a description of the design and the properties of the space-filling fractal grid and describes the burner, the flames studied and the diagnostics used. The results are presented in section \ref{results} and a summary, conclusions and a future outlook are given in section \ref{conclusions}.

\section{Experiment} \label{experiment}

\subsection{The fractal grid}
\begin{figure}[!htp]
\centering
\begin{tabular}{c}
\includegraphics[width=6.7cm]{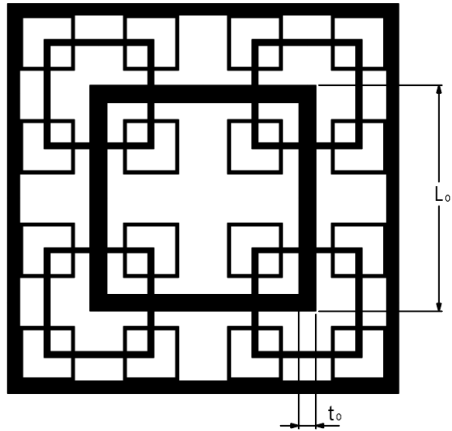}
\end{tabular}
\caption{An image of the fractal grid used in the measurements with relevant geometric quantities.}
\label{fractal_grid}
\end{figure}

The fractal grid, of the same design as in \cite{vassilicos_hurst_fractal_2007}, consists of a main geometric pattern that is repeated at smaller scales and, as the scale decreases, the number of repeated patterns increases. In the grid used in the present measurements the main pattern is a square, whose bars have length $L_0=\unit[36.8]{mm}$ and width $t_0=\unit[2.70]{mm}$. At each successive iteration of the main pattern there are 4 times as many squares as in the previous iteration and the length and width of the grid bars change as $L_j=R_L^jL_0$ and $t_j=R_t^jt_0$, respectively, where $j=0,1,\ldots,N-1$ and $N$ is the number of iterations. The present fractal grid has $R_L=1/2$, $R_t=0.4$ and $N=3$ and is shown in Fig. \ref{fractal_grid}.

The blockage ratio, $\sigma$, of the fractal grid is defined as the ratio of the area covered by the grid to the area of the duct and it is $\sigma=0.22$. An effective mesh size is defined as

\begin{equation}
M_{eff}=\frac{4d^2}{P}\sqrt{1-\sigma}
\label{effective_grid_size}
\end{equation}

\noindent where $d$ is the burner's side (introduced in the following subsection) and $P$ is the perimeter of the fractal grid; here $M_{eff}=\unit[13]{mm}$. The effective mesh size formula when applied to a square mesh grid returns this grid's mesh size.

\begin{figure}[!htp]
\centering
\begin{tabular}{c}
\includegraphics[width=6.7cm]{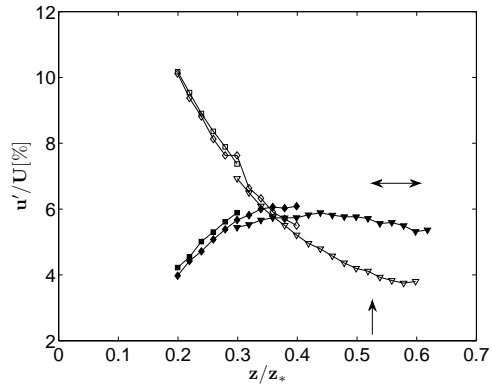}
\end{tabular}
\caption{The decay of the turbulent fluctuations, $u'/U$, with downstream distance. Open symbols correspond to the square grid and filled symbols correspond to the fractal grid. In this plot the point $z=0$ corresponds to the axial position of the grid. The normalization of the downstream distance is with $z_*$, as introduced in the text. The vertical arrow shows the position of the stabilizing rod and the horizontal arrow shows the axial extent of the OH LIF measurements. The conditions in this plot correspond to a bulk velocity $U=\unitfrac[4.3]{m}{s}$, slightly lower than the bulk velocity at flame conditions.}
\label{turbulence_decay}
\end{figure}

As mentioned in the introduction, one of the defining characteristics of this particular design of fractal grids is the difference in the decay of the turbulence intensity with downstream distance from the grid, as compared to a standard square mesh grid. In a standard grid, after a few mesh lengths downstream, the turbulent fluctuations decay following a power law \cite{corrsin_contraction_1966} of the form $\left<u'^2\right>\sim z^{-n}$, where $\left<u'^2\right>$ is the variance of the velocity fluctuations, $z$ is a normalized downstream distance and angle brackets denote averaging. In contrast, it has been shown \cite{vassilicos_hurst_fractal_2007,vassilicos_mazellier_fractal_2009} that in the fractal grid used here the turbulence first \textit{increases} up to a peak value at some distance downstream of the grid before starting to decay. The peak of the turbulence intensity was found to occur at a downstream distance $z_{\mathrm{peak}}\approx 0.5 z_*$, where the wake interaction length scale is

\begin{equation}
z_*=\frac{L_0^2}{t_0}
\label{distance_z_star}
\end{equation}

\noindent where, for the present grid, $z_{\mathrm{peak}}=\unit[226]{mm}$. To demonstrate this hot wire anemometry is used to measure the turbulent fluctuations. Figure \ref{turbulence_decay} 

shows the downstream evolution of the centreline turbulence intensity, $u'/U$ (where $U$ is the local mean velocity), for the fractal grid and a square mesh grid having the same blockage ratio and mesh size. The square mesh grid follows a standard power law decay whereas the turbulent fluctuations in the fractal grid increase with downstream distance before following a slow decay. Furthermore, for downstream distances larger than $\sim$7 mesh sizes from the grids, both the Taylor and integral length scales are continuously larger for the fractal grid generated turbulence. Similarly to \cite{vassilicos_hurst_fractal_2007,vassilicos_mazellier_fractal_2009} (also for grids with N=3), the Taylor length scale is practically constant with downstream distance, whereas the integral length scale increases very slowly. So, the Taylor-based and turbulent (based on the integral length scale) Reynolds numbers have downstream evolutions of similar form to the evolution of the turbulence intensity. Finally, for positions at the centreline and off-centreline, the power spectra of the velocity fluctuations show a broad continuous power-law scaling region with, as expected, larger spectral densities for the fractal grid turbulence.

\begin{figure}[!htp]
\centering
\begin{tabular}{c}
\includegraphics[width=6.7cm]{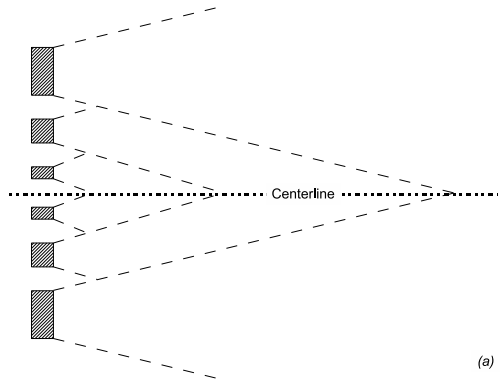}
\end{tabular}
\caption{A schematic explanation of the interaction of the wakes generated from the fractal grid. Reproduced with permissions from \cite{vassilicos_mazellier_fractal_2009}.}
\label{wake_interaction}
\end{figure}

The observed difference in the downstream behavior of the fractal grid turbulence, as compared to the square mesh grid, has been explained in \cite{vassilicos_mazellier_fractal_2009}  by considering the interaction between the wakes generated from the fractal grid bars, each bar having a different size; schematically, this is shown in Fig. \ref{wake_interaction}. 

The smaller wakes reach their peak turbulence intensities closer to the grid. They mix and would decay if it wasn't for the next size-iteration of wakes which reach turbulence intensities further down and thereby help the turbulence to build up to higher intensities further away from the grid rather than decay. This process continues for as many fractal size-iterations as are present on the grid with the result of obtaining, without much pressure drop, a high turbulence intensity with good profile homogeneity at a distance $z_*$ from the grid, \cite{vassilicos_seoud_fractal_2007, vassilicos_mazellier_fractal_2009}. This distance, $z_*$, is given by Eq. \ref{distance_z_star}, as explained by the following argument \cite{vassilicos_mazellier_fractal_2009}. The wake width, $s$, scales as $s\sim \sqrt{tz}$, where $t$ and $z$ are the bar width and the downstream distance, respectively. Taking into account that the largest bar (with width $t_0$ - placed at the furthest distance from the centreline) will create the largest wake, which will contribute more to the turbulent intensity, and assuming that this wake will reach the centreline at $s\sim L_0$, we can arrive at the distance $z_*$ introduced earlier by equating these scalings to get $L_0\sim\sqrt{t_0z_*}$.

It has been shown in \cite{vassilicos_mazellier_fractal_2009} that the mean profile structure of the fractal-generated flow does not change with mean flow velocity and, in particular, that $z_*$ is independent of the mean velocity. Increasing the number $N$ of fractal iterations improves the homogeneity of mean profiles at distances beyond $z_*$. Increasing the overall combustor size and, therefore, the overall fractal grid size makes it easier to use a large number of iterations and therefore works favorably. As shown in \cite{vassilicos_hurst_fractal_2007, vassilicos_mazellier_fractal_2009} who did fluid dynamic experiments in two wind tunnels of different sizes, scaling up does not change Eq. \ref{distance_z_star} even though it obviously changes $z_*$, $L_0$ and $t_0$.

\subsection{The burner}
The experimental burner consists of 2 square ducts, each of inner side $d=\unit[62]{mm}$, oriented vertically upwards. The upstream duct is $L=\unit[500]{mm}$ long and the mixture of fuel and air is injected through 4 inlets at the bottom of this duct. A mixing region of \unit[200]{mm} is followed by a section that comprises a perforated plate with \unit[4]{mm} holes and a mesh screen, which are used to break the large scale structures formed by the 4 incoming jets. After a settling region of \unit[150]{mm}, a second flow conditioning section is placed, with a \unit[50]{mm} long honeycomb with \unit[4]{mm} holes followed by a mesh screen, in order to straighten and laminarize the flow and produce a uniform flowing stream of fuel and air. The turbulence intensity at the position where the grids are placed is around 1\%.

\begin{figure}[!htp]
\centering
\begin{tabular}{c}
\includegraphics[width=6.7cm]{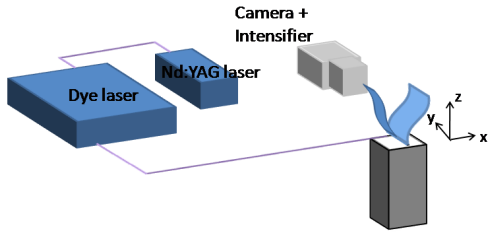}
\end{tabular}
\caption{A sketch of the experimental setup and the coordinate system.}
\label{experimental_setup}
\end{figure}

Turbulence generating grids are placed \unit[100]{mm} downstream of the final mesh screen, exactly at the position where the second, downstream, duct is connected to the upstream one. In the present experiments the length of the downstream duct is \unit[200]{mm}. A turbulent, premixed, V-shaped flame is stabilized on a \unit[2.5]{mm} diameter rod, which is placed \unit[50]{mm} downstream of the exit plane of the burner. The mesh size and the solidity of the standard square mesh grid are the same as the effective mesh size and solidity of the fractal grid. A general view of the burner, the laser system and the coordinate system is shown in Fig. \ref{experimental_setup}.

\subsection{The flames}
\begin{figure}[!htp]
\centering
\begin{tabular}{c}
\includegraphics[width=6.7cm]{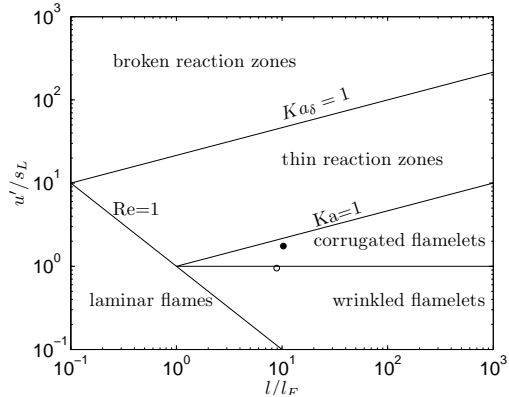}
\end{tabular}
\caption{The position of the two flames in the Borghi diagram. $\circ$, square grid flame; $\bullet$, fractal grid flame. $Ka_\delta=\delta^2Ka$, where $\delta=0.1$ is the thickness of the inner layer compared to the flame thickness, \cite{peters_book}.}
\label{borghi_diagram}
\end{figure}

Two lean ($\phi=0.7$ - $\phi$ is the equivalence ratio), CH$_4$ flames are measured in order to present the differences between the two grids. The air and fuel flow rates are the same for both flames, with a bulk mixture velocity $U=\unitfrac[4.6]{m}{s}$ and, at the position of the stabilizing rod, turbulence intensities, $u'/U$, 4.1\% and 7.6\%, for the square and fractal grids, respectively. In case the mean centerline velocity is used instead of the bulk velocity to normalize the turbulence intensity (see the Results section for an explanation) the corresponding turbulence intensities are 4.1\% and 6\%, respectively. Relevant parameters for both flames are shown in Table \ref{table_parameters} and the position of each flame in the Borghi diagram is shown in Fig. \ref{borghi_diagram}.

\begin{table*}
\centering
\begin{tabular}{cccccccc}
grid & $u'[\unitfrac{m}{s}]$ & $u'/s_L$ & $l[\unit{mm}]$ & $R_t$ & $l/l_F$ & $\alpha [^\circ]$ & $s_T [\unitfrac{m}{s}]$ \\ \hline
square  & 0.19 & 0.95 & 6.4 & 76 & 8.9  & 8  & 0.64 \\
fractal & 0.35 & 1.75 & 7.4 & 145 & 10.3 & 12 & 0.96 \\ \hline
\end{tabular}
\caption{The flame parameters for both the square and the fractal grid. Both flames have the same equivalence ratio, $\phi=0.7$, hence the same laminar flame speed, $s_L=\unitfrac[0.2]{m}{s}$, and flame thickness, $l_F=\unit[0.72]{mm}$. The flame angle is $\alpha$, the turbulent flame speed is calculated as $s_T=U\sin{\alpha}$ and $U=\unitfrac[4.6]{m}{s}$ is the area averaged velocity. The turbulent Reynolds number, $R_t$, is based on the standard deviation of the velocity fluctuations and the integral length scale, $l$.}
\label{table_parameters}
\end{table*}

This plot shows that for the same heat release rate and the same downstream distance, the fractal grid creates more intense turbulence and changes the combustion regime of the flame. The fractal grid flame is placed in the turbulent region of the diagram, whereas the square grid flame is at the borderline between the wrinkled flamelets and the corrugated flamelets regimes. This change in the relevant position of the two flames in the regime diagram implies that turbulence will play a more prominent role in the case of the fractal flame. In order to investigate this, high-speed laser-induced fluorescence imaging of OH has been applied in both flames.

\subsection{High speed OH fluorescence imaging}
The principle of planar laser induced fluorescence measurements can be found in \cite{eckbreth}. A high-speed frequency doubled Nd:YAG laser (Edgewave Innoslab IS8II-DE) laser is used to pump a narrowband, frequency doubled dye laser (Sirah Allegro) that generates around \unit[0.16]{mJ} per pulse at \unit[5]{kHz} repetition rate. Owing to the rather low pulse energies at this high repetition rate, the laser is tuned to excite the strong OH Q1(6) transition in the (1,0) band of the OH A-X system near \unit[283]{nm}. The beam is formed into a light sheet resulting in laser irradiance far below saturation giving a linear dependence of the signal on laser pulse energy. The fluorescence from the (1,1) and (0,0) bands of OH is collected between 305 to 320 nm using a high-speed CMOS camera (LaVision HighSpeedStar 6) coupled to a two-stage high-speed intensified relay optics (LaVision Highspeed IRO). A WG295 Schott glass filter is mounted in front of the f=105, f/2.8 UV camera lens to eliminate any scattered laser light and a UG11 filter is used to suppress flame luminescence and PAH fluorescence. The intensifier gate width is adjusted to \unit[100]{ns} and the full extent of the array, 1024$*$1024 pixels$^2$, is used.

The thickness of the laser sheet at the measurement location is estimated to be \unit[0.2]{mm} and the field of view of the camera is \unit[30$*$30]{mm$^2$} giving a magnification of \unitfrac[0.03]{mm}{pixel}. The integral time scales of the flow are \unit[1.50]{ms} and \unit[1.45]{ms} for the square and the fractal grids, respectively, so that the measurement duration was longer than 600 integral time scales (alternatively, the large eddy turnover time for the square and the fractal grid is \unit[34]{ms} and \unit[21]{ms}, respectively, so the measurements were, at least, 30 turn over times long).

\subsection{Image processing and experimental uncertainties}
The raw OH fluorescent images were processed, using the method developed in \cite{cpiv_beyrau_2007}, in order to extract the instantaneous distributions of the progress variable. The progress variable is a non-dimensional number having the value 0 at the reactants and the value 1 at the products and is defined either as a normalized temperature or product mass fraction \cite{peters_book}. The flame surface density (here: the area of the flame per unit volume), the flame angle and the flame brush thickness are also calculated from these data.

The uncertainty in the equivalence ratio and in the mixture velocity result from the uncertainty in the calibrated flow meters used, which have an accuracy of 2.5\% of the full scale. Therefore, the uncertainties in the equivalence ratio and the bulk mixture velocity are $\pm$\unit[0.025]{} and $\pm$\unitfrac[0.08]{m}{s}, respectively. The statistical uncertainty in the measured standard deviation of the velocity fluctuations is $\pm\unitfrac[8\cdot10^{-4}]{m}{s}$ or $\pm\unitfrac[16\cdot10^{-4}]{m}{s}$, for the square and the fractal grids, respectively. The laminar burning velocity is calculated from a third order polynomial fit to the data of \cite{laminar_flame_speeds_CH4}, as $s_L=\unitfrac[0.2]{m}{s}$ and the thermal flame thickness, $l_F=\unit[0.72]{mm}$, is estimated from the data of \cite{laminar_CH4H2_renou_2008}.

\section{Results} \label{results}

\begin{figure}[!htp]
\centering
\begin{tabular}{c}
\includegraphics[width=6.7cm]{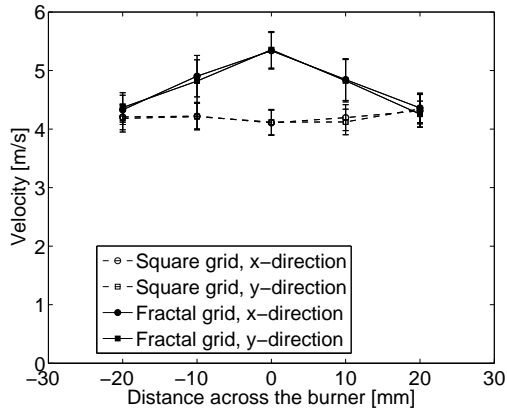}
\end{tabular}
\caption{The velocity profile across the burner, at the burner exit, for both grids (the error bars correspond to one standard deviation of the velocity fluctuations). The conditions in this plot correspond to a bulk velocity $U=\unitfrac[4.3]{m}{s}$, slightly lower than the bulk velocity at flame conditions.}
\label{velocity_profile_cross}
\end{figure}

Other than the downstream evolution of the turbulence intensity, another difference between the square and the present fractal grid is the nozzle exit velocity profiles in Fig. \ref{velocity_profile_cross}, which shows the velocity across the burner at the exit plane, for both grids; the error bars correspond to one standard deviation of the velocity fluctuations at each position. 

The velocity profiles are only presented for the central part of the burner, extending \unit[40]{mm} across, thus excluding the boundary layers and correspond to a bulk velocity $U=\unitfrac[4.3]{m}{s}$, which is slightly lower than the mixture bulk velocity at the flame conditions. In both cases, the profiles show that the flow is symmetric and in the square grid the flow is, also, uniform across the burner. In contrast, the fractal grid imposes a velocity distribution across the burner with higher velocities at the centreline; it should be noted, though, that within the current experimental setup the pressure drop is the same for both grids, so integrating the full velocity profile (including the boundary layers), gives the same area averaged flow rate for both grids. The reason for this inhomogeneity in the case of the fractal grid is, probably, the limited number of fractal iterations. For fractal grids of the same design it has been shown \cite{vassilicos_hurst_fractal_2007,vassilicos_mazellier_fractal_2009,vassilicos_seoud_fractal_2007} that the mean and turbulent velocity profiles across the mean flow direction become more and more homogeneous as the number of iterations increases and are, practically, homogeneous beyond $z_{\mathrm{peak}}$.

\begin{figure*}
\centering
\begin{tabular}{cc}
\includegraphics[width=7cm]{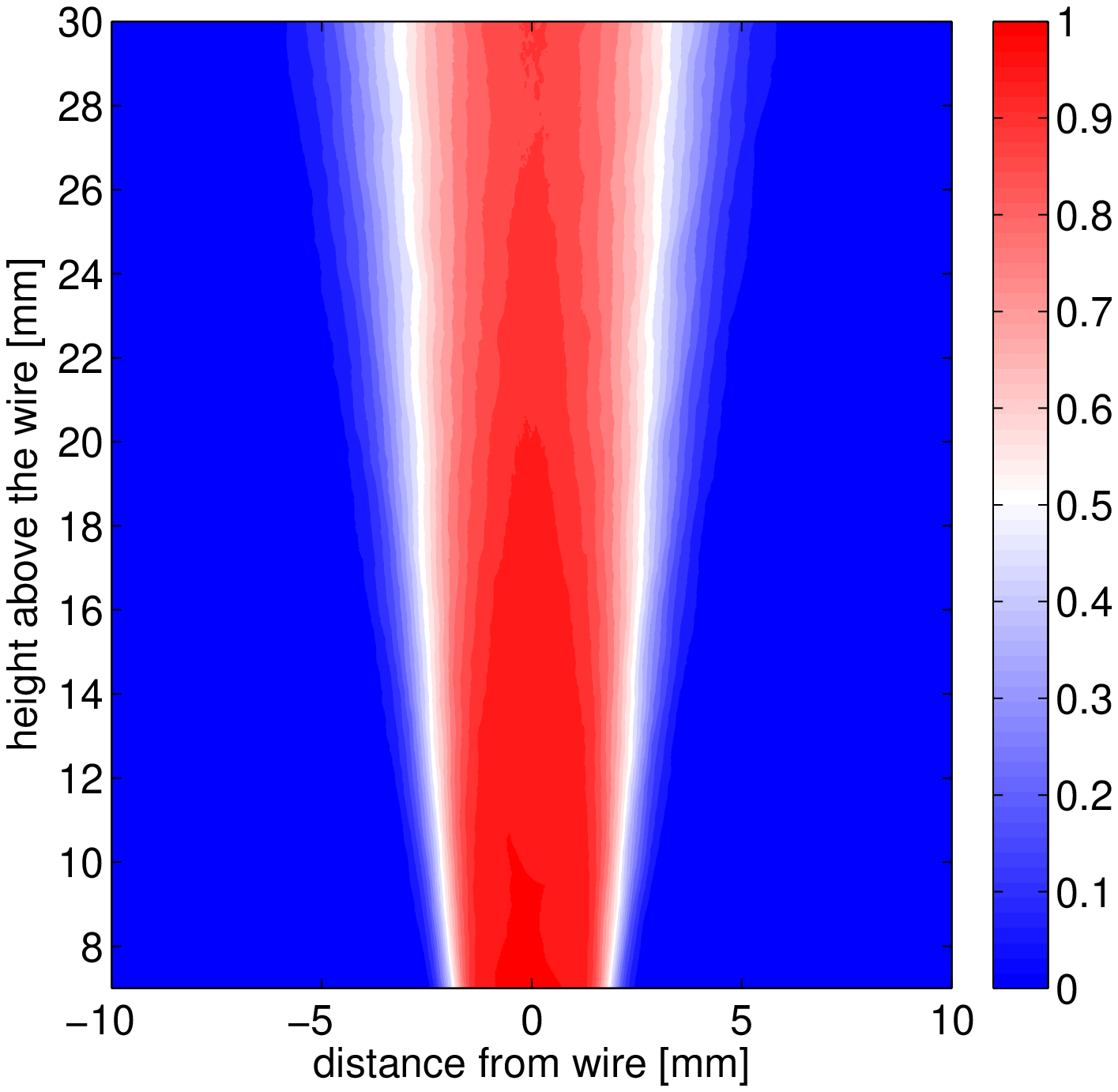} &
\includegraphics[width=7cm]{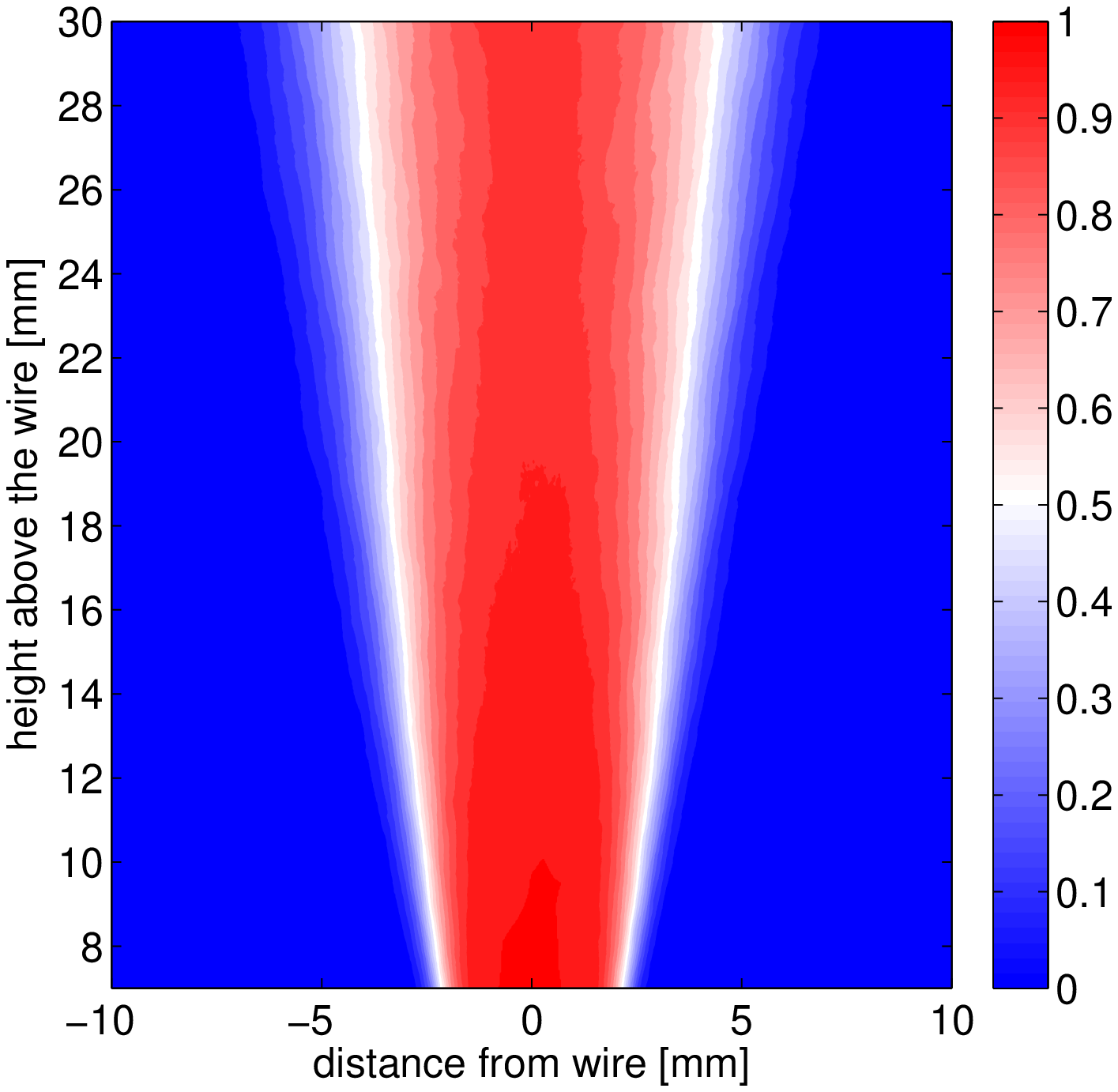}
\end{tabular}
\caption{The mean progress variable distributions for the square grid (left) and the fractal grid (right) flames.}
\label{flame_brush}
\end{figure*}

Figure \ref{flame_brush} shows the progress variable averaged over all the images, for both flames. 

The distribution of the fractal grid flame leans more to the incoming gases than the corresponding square grid flame, implying a larger turbulent flame speed. In fact, the flame angles measured using the $\overline{c}=0.5$ contour line ($c$ being the progress variable) are \unit[8]{$^\circ$} and \unit[12]{$^\circ$} for the square and the fractal grid flames, respectively. Consequently, the turbulent burning velocities (shown in Table \ref{table_parameters}) are different for the two flames, given that the bulk mixture velocity is the same. This should be expected in view of the difference in turbulent fluctuations between the two flames, but it demonstrates the effectiveness of the fractal grid to create more rigorous burning at a given downstream location.

\begin{figure}[!htp]
\centering
\begin{tabular}{c}
\includegraphics[width=6.7cm]{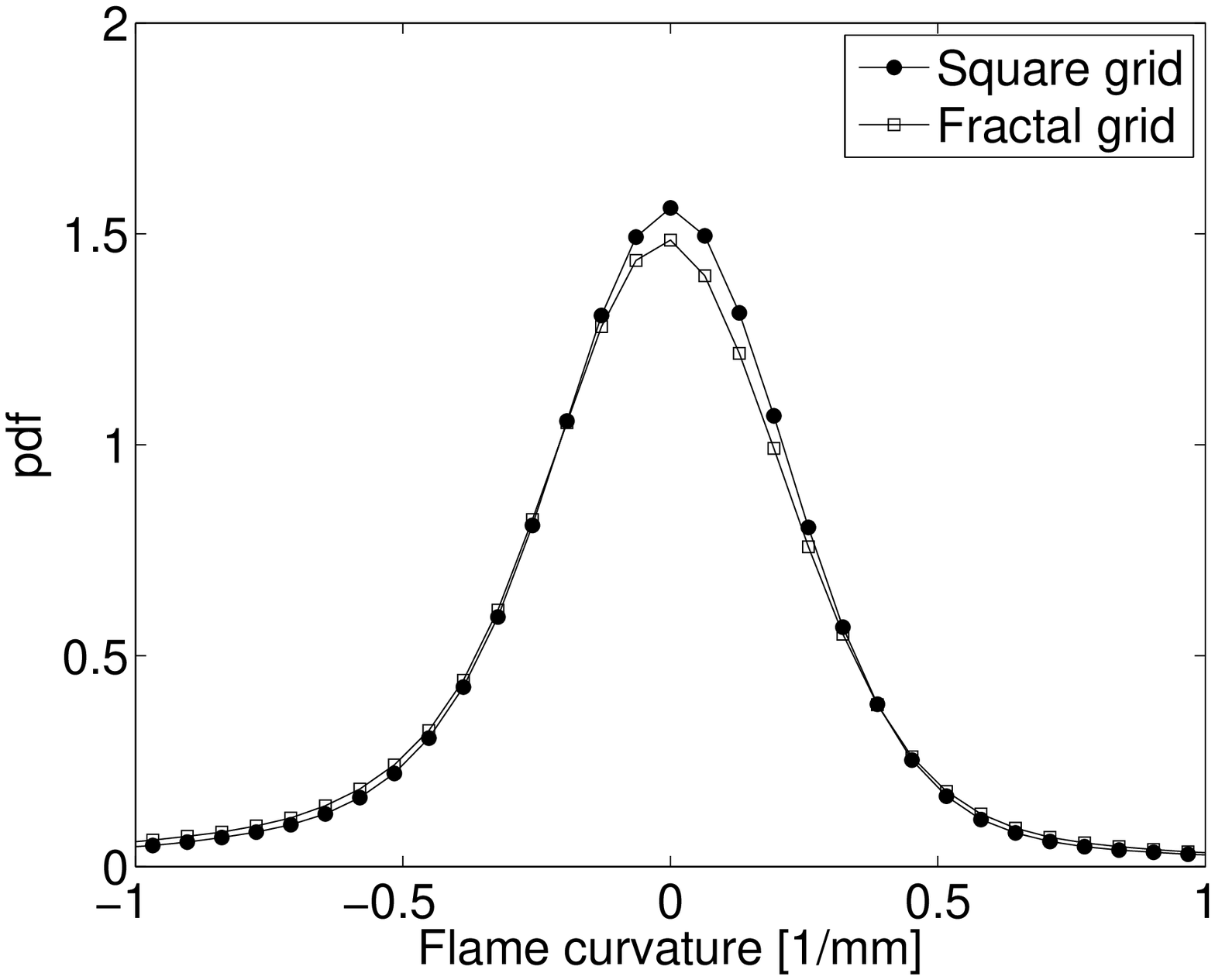} \\
\includegraphics[width=6.7cm]{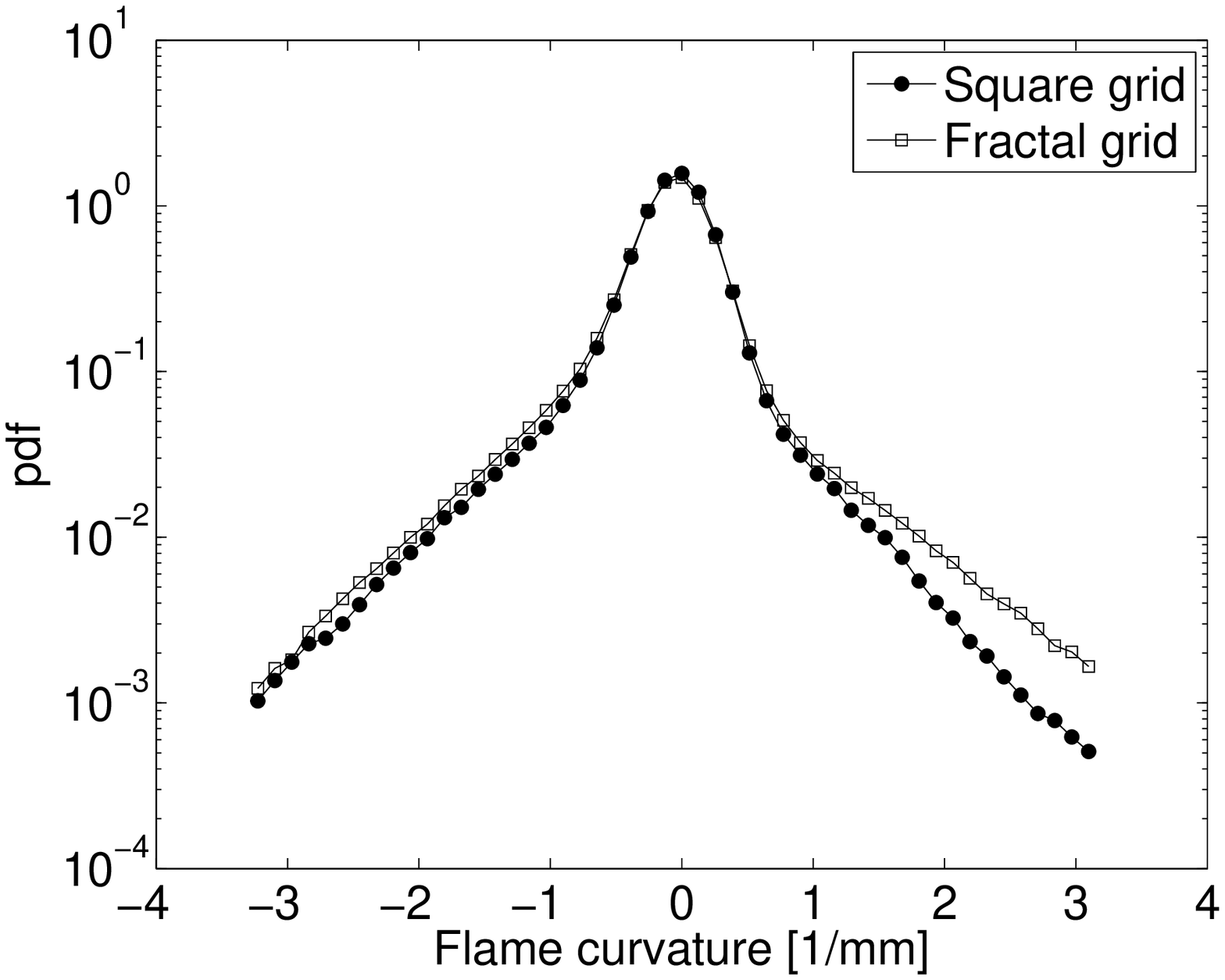}
\end{tabular}
\caption{The probability density functions of curvature for both flames. The semi-logarithmic plot at the bottom demonstrates the large differences at large curvatures between the two flames, whereas the linear plot at the top shows that the square grid flame has a larger occurrence of small and near zero values of curvature.}
\label{curvature_pdf}
\end{figure}

With increased turbulent fluctuations, compared to the laminar flame speed, the flame will become more and more corrugated. The square grid flame sits at the borderline between the wrinkled and corrugated flamelets regimes, whereas the fractal grid flame is in the corrugated flamelets regime, nearer the thin reaction zones regime. Consequently, for the fractal grid flame, the motion of turbulent eddies will dominate over the movement of the flame front with the laminar burning velocity, these two effects being of comparable magnitude at the square grid flame. Quantitatively this difference is shown in Fig. \ref{curvature_pdf}, which plots the probability density functions of the curvature for the two flames (positive curvatures denote flame front "excursions" towards the reactants). 

The pdf are shown in both linear and semi-logarithmic plots, in order to focus at small and large values of the curvature. The distributions are symmetric and similar to each other near the mean value. The tails of both distributions are more pronounced than a normal distribution (with the same mean and standard deviation) and become increasingly asymmetric at larger curvature values, more so for the square grid flame. The pdf of the square grid flame is higher and slightly wider near the mean value and it falls off quicker. This is more evident in the semi-logarithmic plot that shows the much fatter tails of the distribution of the fractal grid flame, especially at large positive values of the curvature.

\begin{figure}[!htp]
\centering
\begin{tabular}{c}
\includegraphics[width=6.7cm]{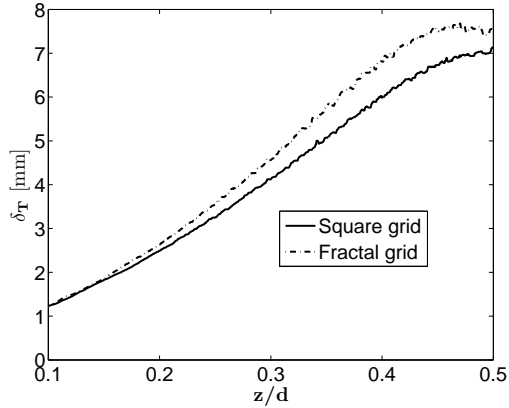}
\end{tabular}
\caption{The downstream evolution of the flame brush thickness, $\delta_T$.}
\label{brush_thickness_downstream}
\end{figure}

The downstream dependence of the flame brush thickness, $\delta_T$, is shown in Fig. \ref{brush_thickness_downstream}. The flame brush thickness was computed from the flame surface density distribution (derived from the progress variable, but not shown here, using a procedure outlined in \cite{cpiv_beyrau_2007}) by fitting the sum of two gaussian functions to the transverse profile at each downstream distance and calculating the flame brush thickness as the average of the standard deviations of the two gaussian functions (each gaussian corresponds to one "leg" of the V-flame). 

The flame brush thickness is an indication of the average movement of the flame due to corrugations induced by the turbulent flow field and, in the present flames, it is consistently thicker for the fractal grid flame (for distances further downstream of the stabilizing rod) manifesting the higher turbulence intensities in this flame. In fact, the Taylor and integral length scales are, also, larger in the fractal grid flame, so they induce a larger motion of flame. For both flames, in the current combustion regimes, large scale effects, depending on turbulence intensities and integral length scales, are expected to be more important. Smaller scale wrinkling of the flame would be more pronounced as one reaches the $Ka=1$ line in the combustion regime diagram, where the flame could thicken further due to corrugations caused by the smallest eddies in the flow penetrating the preheat zone of the flame. However, no attempt is currently made to identify such effects, which requires comparing the instantaneous thickness of the preheat zone with the smallest flow scales.

\begin{figure}[!htp]
\centering
\begin{tabular}{c}
\includegraphics[width=6.7cm]{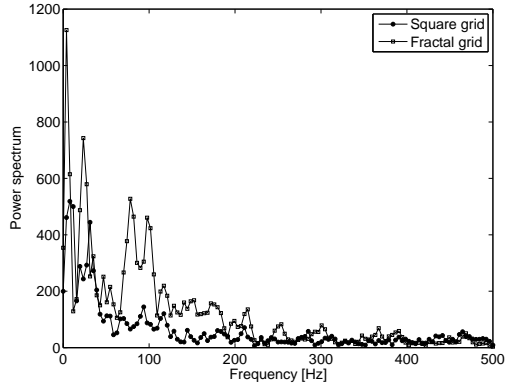}
\end{tabular}
\caption{The power spectrum of the flame length time series.}
\label{fsd_spectrum}
\end{figure}

The dynamics of the flame are, initially, explored by examining the time series of the flame length, as deduced from the instantaneous flame front. The power spectrum is shown in Fig. \ref{fsd_spectrum}, for both grids. 

As a reference, the shedding frequency from the bars of the fractal grid is calculated from the Strouhal number as $St=ft/U$, where $St$ is the Strouhal number and $f$ is the shedding frequency. Assuming a Strouhal number $\approx0.12$, the shedding frequency from the largest bar is $f_0\approx\unit[200]{Hz}$; in spite of that the various peaks are generally at lower frequencies than the shedding frequencies of the various bars. However, the peaks in the fractal grid case are generally of larger magnitude, implying that the fluctuations associated with the flame area are, also, of larger magnitude in the fractal grid flame. Another potentially relevant frequency that could be related to this spectrum is the transverse "meeting" frequency of the wakes from the various bars, but this frequency wasn't estimated.

\subsection{Discussion}
In the following we, briefly, assess the effect of fractal-grid generated turbulence on the turbulent burning velocity. In the corrugated flamelet regime, where large flow scales are more important than small scale turbulence, we can write \cite{peters_book}

\begin{equation}
\frac{s_T}{s_L}\sim\frac{A_T}{A_L}
\end{equation}

\noindent where $A_T$ and $A_L$ are the flame surface area in the turbulent flame and $A_L$ is the flame surface area of a laminar flame. With some generality, one can consider that the flame surface ratio $A_T / A_L$ is set by the ratio of a large (outer) length scale $L_{ref}$ (dependent on geometry) to a small (inner) length scale $l$. Our argument in this section in only indicative, and we therefore adopt a simple way to model this dependence as a power law

\begin{equation}
\frac{A_T}{A_L}\sim\left(\frac{L_{ref}}{l}\right)^{D-2}
\label{gouldin_fractal_model}
\end{equation}

\noindent where $D-2 > 0$. If the flame was a fractal object then the exponent $D$ would be its fractal dimension as proposed by \cite{gouldin_fractal_model_1987}, but we do not need to rely on such an assumption here. It is reasonable to expect the ratio of $l$ to the integral length scale $L$ to be an increasing function of the ratio $s_{L}/u'$. For the sake of argument, we assume

\begin{equation}
\frac{l}{L} \sim \left(\frac{s_{L}}{u'}\right)^{h}
%\eqno(5)
\end{equation}

\noindent where the exponent $h$ is positive. In \cite{peters_flamelet_concepts_1986} and \cite{kerstein_linear_eddy_model_1988} it is argued that $h=3$, but our argument does not need this value. From the above two relations

\begin{equation}
\frac{s_{T}}{s_{L}} \sim \left(\frac{u'}{s_{L}}\right)^{h(D-2)} \left(\frac{L_{ref}}{L}\right)^{D-2}.
%\eqno(6)
\end{equation}

\noindent The above scaling is valid locally for a flamelet and we assume that we can relate the flame angle calculated from Fig. \ref{flame_brush} with the local flamelet structure by integrating for the downstream dependence of $u'$ and $l$. We can take for the square grid $u'\sim u_0'\left(z/L_{ref}\right)^{-n/2}$ and $l/L_{ref}\sim\left(z/L_{ref}\right)^{1-n/2}$, \cite{corrsin_contraction_1966}, and for the fractal grid $u'\sim u_0'e^{-2z/z_*}$ (valid for distances larger than $z_*$, see \cite{vassilicos_mazellier_fractal_2009}) and an approximately constant integral length scale $l\sim z_*$ ($u_0'$ is the incoming velocity rms). We choose $h=3$ and $D=2+1/3$, mainly for convenience and to recover Damk\"{o}hler's scaling, without this choice affecting the qualitative discussion and we obtain

\begin{eqnarray}
  S_{T, normal}  & \sim & u_0' \\
  S_{T, fractal} & \sim & u_0'\left(z_*/L_{ref}\right)^{2/3}
\end{eqnarray}

\noindent where $S_T=\int s_T(z)dz/L_{ref}$. The outer scale $L_{ref}$ may be expected to scale with the integral scale $L$, which, in turn, has been shown to scale with $L_{0}$, \cite{vassilicos_hurst_fractal_2007, vassilicos_seoud_fractal_2007, vassilicos_mazellier_fractal_2009}. Hence, from Eq. \ref{distance_z_star}, $z_*/L_{ref} \sim L_{0}/t_{0}$. These expressions attest to an important qualitative difference between the effect on the flame, at the corrugated flamelets regime, from the square mesh and fractal grids, \textit{i.e.} that the fractal grid generated turbulence imposes, also, a length scale dependence on the turbulent burning velocity, which can be manipulated by modifying the lengths and widths of the bars of the fractal grid.

\section{Conclusions} \label{conclusions}

The presented results demonstrate the effect and the value of the fractal grid as a turbulence generator in premixed turbulent combustion. It is shown that by using the fractal grid more intense turbulence can be generated at a given downstream distance compared to a normal square mesh grid and, as a consequence, the turbulent fluctuations and the turbulent flame speed are increased. In fact the flame angle and the turbulent flame speed increased by 50\% by using the fractal grid. In the future, this will prove beneficial, \textit{e.g.} for extending the lean stability limit for a given heat release rate. Analysis of the flame quantities reveals that both the curvature and the flame brush thickness of the fractal grid flame are more pronounced, in the sense that the flame presents larger corrugations and more intense burning than the normal grid flame.

\paragraph{Future Outlook}
The frequency content of the flame length time series shows some characteristic frequencies which have, however, not been analyzed in detail yet. We believe that further analysis of the time resolved data will allow the identification of dynamic effects on flame development. For example, Proper Orthogonal Decomposition (POD) analysis will help identify these dynamics, and simultaneous high-speed PIV will provide both spatial flow structure information (that is practically non-existent for the present grid design) and correlations between turbulence and combustion quantities. Extinction events, which are also captured within the present measurements and their effect on the flame will be investigated. Finally, the adverse effect of the mean flow inhomogeneity will be addressed, and can be ameliorated, by altering the design of the fractal grids.

\section{Acknowledgements}
Financial support from Imperial Innovations Group plc for part of the project is gratefully acknowledged. We wish to thank Ms Monica Luegmair for her help with the high-speed measurements.



%% References

\section{References}
\bibliographystyle{pci}
\bibliography{References}

\begin{thebibliography}{10}
\expandafter\ifx\csname url\endcsname\relax
  \def\url#1{\texttt{#1}}\fi
\expandafter\ifx\csname urlprefix\endcsname\relax\def\urlprefix{URL }\fi

\bibitem{lean_combustion_gas_turbines_mcdonnel_2008}
V.~McDonnel, in: D.~Dunn-Rankin (Ed.), Lean combustion {--} {T}echnology and
  control, Academic Press, 2008.

\bibitem{driscoll_mongia_flashback_2009}
S.~K. Dhanuka, J.~E. Temme, J.~F. Driscoll, H.~C. Mongia, {\em Proceedings of
  the Combustion Institute\/} 32~(2) (2009) 2901 -- 2908.

\bibitem{candel_flashback_2004}
D.~Bernier, F.~Lacas, S.~Candel, {\em Journal of Propulsion and Power\/}
  20~(4).

\bibitem{bradley_turbulent_velocity_1981}
R.~G. Abdel-Gayed, D.~Bradley, M.~Lawes, {\em Proceedings of the Royal Society
  of London. A. Mathematical and Physical Sciences\/} 414~(4) (1987) 389--413.

\bibitem{cheng_flame_grid_turbulence_1981}
R.~Bill~Jr., I.~Namer, L.~Talbot, R.~Cheng, F.~Robben.

\bibitem{vassilicos_fractal_prl_2010}
R.~Stressing, J.~Peinke, R.~E. Seoud, J.~C. Vassilicos, {\em Physical Review
  Letters\/} accepted for publication.

\bibitem{vassilicos_fractal_mixing_effective_2009}
C.~Coffey, G.~Hunt, R.~Seoud, J.~C. Vassilicos, {\em Experimental Thermal and
  Fluid Science,\/} submitted for publication.

\bibitem{vassilicos_hurst_fractal_2007}
D.~Hurst, J.~C. Vassilicos, {\em Physics of Fluids\/} 19~(3).

\bibitem{corrsin_contraction_1966}
G.~Comte-Bellot, S.~Corrsin, {\em Journal of Fluid Mechanics Digital Archive\/}
  25~(04) (1966) 657--682.

\bibitem{vassilicos_mazellier_fractal_2009}
N.~Mazellier, J.~C. Vassilicos, {\em Physics of Fluids\/} accepted for
  publication.

\bibitem{vassilicos_seoud_fractal_2007}
R.~E. Seoud, J.~C. Vassilicos, {\em Physics of Fluids\/} 19~(10).

\bibitem{eckbreth}
A.~C. Eckbreth, {\em Laser diagnostics for combustion temperature and
  species\/}, Gordon and Breach Publishers, 1996.

\bibitem{cpiv_beyrau_2007}
S.~Pfadler, F.~Beyrau, A.Leipertz, {\em Optics Express\/} 15 (2007) 15444 --
  15456.

\bibitem{peters_book}
N.~Peters, {\em Turbulent combustion\/}, Cambridge University Press, 2000.

\bibitem{laminar_flame_speeds_CH4}
X.~J. Gu, M.~Z. Haq, M.~Lawes, R.~Woolley, {\em Combustion and Flame\/}
  121~(1-2) (2000) 41 -- 58.

\bibitem{laminar_CH4H2_renou_2008}
Y.~Lafay, B.~Renou, G.~Cabot, M.~Boukhalfa, {\em Combustion and Flame\/}
  153~(4) (2008) 540 -- 561.

\bibitem{gouldin_fractal_model_1987}
F.~Gouldin, {\em Combustion and Flame\/} 68~(3) (1987) 249 -- 266.

\bibitem{peters_flamelet_concepts_1986}
N.~Peters, {\em Proceedings of the Combustion Institute\/} 21 (1986) 1231 --
  1250.

\bibitem{kerstein_linear_eddy_model_1988}
A.~R. Kerstein, {\em Combustion Science and Technology\/} 60~(4) (1988) 391 --
  421.

\end{thebibliography}

%%\bibliography{REFERENCES}





% \end{thebibliography}

\end{document}